\documentclass{acm_proc_article-sp}

\newif\ifpdf
\ifx\pdfoutput\undefined
\pdffalse 
\else
\pdfoutput=1 
\pdftrue
\fi

\def\A{{\tt A}}
\def\C{{\tt C}}
\def\G{{\tt G}}
\def\T{{\tt T}}
\def\CpG{{\tt CpG}}
\def\X{{\alpha}}
\def\Y{{\beta}}
\def\Z{{\gamma}}
\def\ZZ{{\delta}}
\def\bX{{\overline{\alpha}}}
\def\bY{{\overline{\beta}}}
\def\bZ{{\overline{\gamma}}}
\def\bZZ{{\overline{\delta}}}

\def\fhat{\hat{f}}
\def\rhohat{\hat{\rho}}

\newcommand\rateQ[2]{Q_{#1#2}}
\newcommand\rateR[2]{R_{#1#2}}

\def\Q{{\bf Q}}
\def\R{{\bf R}}
\def\L{{\cal L}}

\setpagenumber{1}
\begin{document}
%

\title{DNA Sequence Evolution\\
with Neighbor-Dependent Mutation}
%
%

\numberofauthors{3}
%

\author{
%
\alignauthor 
	Peter F. Arndt\\
       \affaddr{Department of Physics}\\
	\affaddr{UC San Diego}\\
	\affaddr{9500 Gilman Drive}\\
	\affaddr{La Jolla, CA 92093}\\
            \email{arndt@physics.ucsd.edu}
\alignauthor Christopher B. Burge\\
	\affaddr{Department of Biology}\\
	\affaddr{MIT}\\ 
	\affaddr{77 Massachusetts Ave.}\\
	\affaddr{Cambridge, MA 02139}\\
       \email{cburge@mit.edu}
\alignauthor Terence Hwa\\
       \affaddr{Department of Physics}\\
	\affaddr{UC San Diego}\\
	\affaddr{9500 Gilman Drive}\\
	\affaddr{La Jolla, CA 92093}\\
            \email{hwa@matisse.ucsd.edu}
}
\date{20 December 2001}
\maketitle
\begin{abstract}
We introduce a model of DNA sequence evolution which can account for
biases in mutation rates that depend on the identity of the neighboring
bases.
An analytic solution for this class of models is developed by adopting 
well-known methods of nonlinear dynamics. Results are presented for
the \CpG-methylation-deamination process which dominates point
substitutions
in vertebrates. The dinucleotide frequencies generated by the model (using
empirically obtained mutation rates) match the overall pattern  observed
in non-coding DNA. A web-based tool has been constructed to
compute single- and dinucleotide frequencies for arbitrary
neighbor-dependent mutation rates.
Also provided is the backward procedure to infer the mutation
rates using maximum likelihood analysis   
given the observed single- and dinucleotide frequencies. 
Reasonable estimates of the 
mutation rates can be obtained very efficiently, using generic 
non-coding DNA sequences as input, after masking out 
long homonucleotide subsequences.
Our method is much more convenient
and versatile to use than the traditional method of deducing
mutation rates by counting mutation
events in carefully chosen sequences.
More generally, our approach provides a more realistic but  
still tractable description of non-coding genomic DNA, and
may be used as a null model for various
sequence analysis applications.

\end{abstract}



\vspace*{1mm}

\section{Introduction}

\vspace*{1mm}

Models of DNA sequence evolution have generally treated sequences as
collections of independently evolving sites\cite{Ya94}.  
However, there is a substantial
amount of evidence from sequence analysis and studies of mutation rates\cite{blake} which
suggests that the identities of neighboring bases can have a strong influence
on the types and rates of mutational events which occur at a given sequence
position.  We briefly review these types of evidence before describing a
model which explicitly accounts for neighbor-dependent mutations.

Biochemical studies in the early 1970s compared the pattern of dinucleotide
odds ratios (dinucleotide frequencies normalized for the base composition)
or ``general designs'' of different genomes and different fractions of genomic
DNA\cite{Ru77,Ru76} and concluded that this pattern is a remarkably stable property of a
genome which is largely preserved in closely related genomes or in different
renaturation rate fractions of the same genome, for example.  Some two decades
later, a significant body of sequence analysis work primarily by Karlin and
coworkers\cite{KaBu95,KM97,Karetal97} has elaborated and
expanded on these observations, showing that the pattern of dinucleotide
relative abundance values (essentially equivalent to the general design)
constitutes a ``genomic signature'' in the sense that it is remarkably constant
across different parts of a genome and is generally similar between related
organisms, but quite different between distantly related organisms.  This
latter finding has led to application of this principle to phylogeny
reconstruction and identification of laterally transferred genes in both
prokaryotic and eukaryotic organisms.  Although quite promising, this line of
work is open to the criticism that there is no underlying theory for why a
genome should possess a particular signature or the mechanism by which
signatures change over long periods of time.  The model described here may help
to provide a theoretical framework for interpreting these and related
observations. 
Our analysis may also contribute to understanding of the
mechanisms of DNA mutation and repair and may help in the development of better methods
for phylogenetic tree construction.

Possible factors relating to the genome signature include effects related to
DNA structure, base stacking and thermodynamics as well as a variety of
mutational factors.  Since the genome signature is particularly pronounced in
non-coding sequences which are typically under much less selective pressure
than coding regions and evolve much more rapidly, it appears likely that
``nonselective'' forces such as biases related to mutation, replication and DNA
repair account for many of the properties of the genome signature, especially
in higher eukaryotic organisms with large genomes containing mostly non-coding
DNA and relatively small effective population sizes.  
One of the most common types of
mutation in vertebrate genomes is \CpG\ methylation followed by deamination and
mutation of \C\G/\C\G\ to \T\G/\C\A\ \cite{CG1,CG2}.  
For example, Wang {\it et al.}\cite{Wanetal98}
found that single-nucleotide polymorphisms occur approximately tenfold
more often at \CpG\ dinucleotides than at other dinucleotides in human genomic
DNA, suggesting that this is by far the most common type of single base
mutation in humans. This is in agreement with the earlier finding
by Batzer {\it et.\ al.}\cite{Batzer} and Hess {\it et.\ al.}\cite{blake}
 that point substitutions occur 10 times more frequently at \CpG\ sites
compared to non-\CpG\ sites.
   Increased mutation rate associated with \CpG\ methylation
can explain at least qualitatively why \CpG\ dinucleotides have consistently
extremely low relative abundance values in all vertebrate nuclear
genomes\cite{KM97}.
This effect was recently exploited by Fryxell and Zuckerkandl\cite{FrZu00}
in their theory of the origin of genomic isochores in mammals.

Here we consider a model for DNA sequence evolution which includes
neighbor-dependent mutation effects. Efficient computational tools are 
constructed
to solve the model for arbitrary user-specified mutation processes and rates.
The main results are quantitative answers to (1) the long term effects of 
neighbor-dependent mutations  on the base
compositional structure of a genome, and (2) the underlying mutation processes
and rates given sequence data with dinucleotide correlations.
 In principle, such a model should be able
to account for the effects of \CpG-methylation induced mutation and several
other known contextual effects on mutation rates.  However, in order to
preserve the tractability of the model, certain other types of mutations which
occur in nature had to be ignored.  For example mutations which change the
lengths of DNA sequences (insertions and deletions such as those caused by
polymerase slippage) or more complex mutations such as inversions or other
large-scale rearrangements are not considered.  Nevertheless, the current model
is a reasonable first approximation to DNA sequence evolution in
the absence of selection.  

This paper is organized as follows: In Sec.\ 2, we introduce our model of
sequence evolution with neighbor-dependent mutation, and define the mutation
rate matrices. In Sec.\ 3, we describe the general scheme used to calculate 
the single and dinucleotide frequencies, and present the pattern of dinucleotide
odds ratio obtained
for a particular example involving the \CpG-methylation-deamination process
described above. We also present the backward analysis, where 
a maximum likelihood approach is used to {\em infer} the mutation 
rates from the observed dinucleotide frequencies within the confines of
our model. A web server is made available for the public 
to perform these
calculations at {\tt http://bioinfo.ucsd.edu/dinucleotides}. In Sec.\ 4, we
list some of a large number of studies made possible by the method described
here. Details of the method are relegated to the Appendix.

\section{The Model}
\vspace*{1mm}
We consider a sequence of $L$ nucleotides $\X_1,\X_2,\dots,\X_L$
with $\X_i \in \{\A,\C,\G,\T\}$. The configuration 
at time $t$ is denoted by $\vec\X(t)=(\X_1(t),\X_2(t),\dots,\X_L(t)\,)$.
There are two types of mutation processes allowed:
(1) mutations of a single nucleotide $\X_i\rightarrow\X_i'$ which occurs
 with a rate $Q_{\X\X'}$ independent of 
its neighbors, and
(2) mutations of a pair of neighboring nucleotides
$\X_i\X_{i+1}\rightarrow\X_i'\X_{i+1}'$ which occurs with a 
rate $R_{\X\Y\X'\Y'}$.
These rates are positive numbers and fixed in time.

We start with an initial random sequence $\vec\X(0)$.
The dynamics of the model is Markovian. At any time $t$, 
the sequence $\vec\alpha(t)$ is updated 
in discrete time steps $\Delta t/L$ according to the following 
update rules:
\begin{enumerate}
\item 
A position $i$ is chosen at random between $1$ and $L$;

\item
The nucleotide $\X_i$ from the sequence $\vec\X(t)$ is 
mutated to $\X_i'$ with 
        \begin{equation}
        {\rm Pr}(\X'_i|\X_i)=
        \left\{ \begin{array}{ll}
                1+ Q_{\X_i\X_i} \cdot \Delta t & \mbox{ for }
\X_i = \X'_i \\
                Q_{\X_i\X'_i} \cdot \Delta t & \mbox{ otherwise}
                \end{array}
        \right.
	\label{rulei}
        \end{equation}
to generate an intermediary sequence  $\vec\X'(t)$;

\item 
A pair of neighboring positions $j$ and $j+1$ is chosen at random 
between $1$ and $L$;

\item
The nucleotides $\X'_j\X'_{j+1}$ from the sequence $\vec\X'(t)$
 are mutated  to $\X''_j\X''_{j+1}$ with
	\begin{eqnarray}
        &&{\rm Pr}(\X''_j\X''_{j+1}|\X'_j\X'_{j+1})=
		\nonumber\\
        &&\quad
		\left\{ \begin{array}{ll}
                1+ R_{\X'_j\X'_{j+1}\X'_j\X'_{j+1}} \cdot \Delta t\\
		\qquad\qquad
		 \mbox{ for }  \X'_j =  \X''_j
                 \mbox{ and }\X'_{j+1} =\X''_{j+1}\\
                R_{\X'_j\X'_{j+1}\X''_j\X''_{j+1}} \cdot \Delta t\\
		\qquad\qquad
                  \mbox{ otherwise}
                \end{array}
        \right.
	\label{ruleii}
        \end{eqnarray} 
to generate the new sequence  $\vec\X(t+\Delta t/L)$. 

\end{enumerate}

To guarantee the conservation of the transition probabilities, the rates
must be chosen such that $\rateQ{\X}{\X}=-\sum'_{\X'}\rateQ{\X}{\X'}$
and $\rateR{\X\Y}{\X\Y}=-\sum'_{\X'\Y'}\rateR{\X\Y}{\X'\Y'}$.
The time increment $\Delta t$ should be chosen such that 
all non-diagonal transition probabilities in Eqs.~(\ref{rulei}) 
and (\ref{ruleii}) are small ($\ll 1$).
Then after $L$ such iterations, on average every base in the sequence
 $\vec\alpha(t)$ is mutated with probability
$(\rateQ{}{} + \rateR{}{}) \cdot \Delta t$, corresponding to a net increment
of time by $\Delta t$.
The above procedure is then repeated
for a long time\footnote{Dynamical aspects of this evolution model 
will be discussed elsewhere\cite{ArHw}.}
 until the {\em stationary state} is reached. Since the
model is ergodic, the stationary state is unique. We denote
the probability to find a configuration $\vec\X$ in the stationary state by 
$P(\vec\X)$.

We refer to rates $\rateQ{\X}{\X'}$  and $\rateR{\X\Y}{\X'\Y'}$ 
collectively as the mutation matrix $\Q$ and $\R$, respectively.
There are a total of 12 independent rates $\rateQ{\X}{\X'}$ 
and 240 independent rates $\rateR{\X\Y}{\X'\Y'}$.
Note that in the special case of neighbor-dependent single
nucleotide mutation, e.g., the \CpG-methylation-deamination process 
described in Sec.~1, $\R$ is given
by a {\em restricted} form with $\rateR{\X\Y}{\X'\Y'}$ non-zero only if
$\X=\X'$ or $\Y=\Y'$.
There are still 96 independent rates under this restriction.
For non-coding DNA sequences, mutations on the complementary DNA strand should
generally occur at the same rate as on the forward strand.
We therefore expect a {\em reverse complementary symmetry} in the rates, i.e.,
$\rateQ{\X}{\Y}=\rateQ{\bX}{\bY}$,
and $\rateR{\X\Y}{\Z\ZZ}=\rateR{\bY\bX}{\bZZ\bZ}$, where
$\bX$ denotes the complement of $\X$, e.g. $\overline{\A}=\T$.
These complementarity conditions reduce the number of independent parameters 
by another factor of two. Note however that for the calculation to be presented below,
it is not necessary to impose any of these conditions on $\Q$ or $\R$.

\section{Methods and Results}
\label{secres}

\vspace{1mm}
The subject of this study is the stationary probability distribution 
$P(\vec\X)$ for the mutation 
processes characterized by the rates $\Q$ and $\R$. If each nucleotide position
evolves independently of each other, i.e., if $\R = 0$, then 
the distribution factorizes, with  $P(\vec\X)
=\prod_i P_0(\X_i)$,
where $P_0(\X)$ is given by the eigenvector corresponding to the largest
eigenvalue of  $\Q$.
However, for $\R \neq 0$,
the bases in the sequence do not
evolve independently and the stationary distribution is difficult to
calculate.
Instead, we focus on the single nucleotide frequencies
(or the ``base composition'') and the dinucleotide frequencies, given by
\begin{equation}
f_\X \equiv \frac1L \sum_{i=1}^{L} 
\sum_{\vec\Y} \delta_{\Y_i\X} P(\vec\Y)
\end{equation}
\begin{equation}
f_{\X\Y} \equiv \frac1{L-1} \sum_{i=1}^{L-1} 
 \sum_{\vec\Z} \delta_{\Z_i\X}\delta_{\Z_{i+1}\Y}  P(\vec\Z),
\label{f2}
\end{equation}
respectively.
An important motivation 
for computing these frequencies is that they are 
easily measured from actual sequence data\cite{KaBu95}
and can thus be used  for quantitative comparison with the output of our
model.

\subsection{Forward analysis: Computation of nucleotide frequencies}

\vspace{1mm}

Our first goal is to compute the frequencies $f_{\X}$ and $f_{\X\Y}$ 
in the stationary state for any set of mutation rates $\Q$ and $\R$.
Exact solution of these quantities is still difficult because in order to 
compute the dinucleotide frequencies, one needs to know the trinucleotide
frequencies $f_{\X\Y\Z}$, etc., ending up with an infinite hierarchy of equations. 
This is a frequently encountered problem in coupled dynamical
systems.  Here we introduce an approximation procedure
called the ``two-cluster
approximation'' which is well-known from nonlinear dynamics\cite{bAKo92}.
This procedure
truncates the hierarchy of equations, expressing the trinucleotide frequencies
as a function of the single and dinucleotide frequencies,
i.e., 
\begin{equation}
f_{\X\Y\Z} = f_{\X\Y}f_{\Y\Z}/f_{\Y}\,, 
\label{approx0}
\end{equation}
and then solves for
the dinucleotide frequencies simultaneously. The procedure gives the
exact solution if the stationary state of the mutation process $\R$
is a first-order Markov chain; it is generally very accurately as long as 
the sequence correlation in the stationary state is short-ranged. Comparison
of solutions obtained using this method with the best numerical estimate
from Monte-Carlo simulation yielded relative disagreement well below the $1\%$ 
level; see below. Thus, for practical purposes, we can regard the cluster approximation
as generating ``exact'' results. 
The advantage of using the cluster approximation
(instead of Monte-Carlo simulation) is that the single and dinucleotide frequencies
can be computed virtually instantaneously
by numerically solving a set of algebraic
equations for {\em arbitrary} mutation matrices $\Q$ and $\R$, i.e., without
any constraints on the rates.
We have
developed a web server at {\tt
http://bioinfo.ucsd.edu/} {\tt dinucleotides} to perform this calculation.

We will present our method here via a specific example, the 
\CpG-methylation-deamination process described in the introduction.
This process is described by a single neighbor-dependent mutation rate,
i.e.
	\begin{equation}
	\rateR{\C\G}{\C\A} = \rateR{\C\G}{\T\G} = r
	\label{eqr}
	\end{equation}
and $\rateR{\X\Y}{\Z\ZZ} = 0$ 
for all other nondiagonal entries. 
To make this example transparent, we also consider a simplified version
of the neighbor-independent mutation rate $\Q$, adopting a single rate $q$ 
for all transitions and another rate $p$ for all transversions, i.e., 
	\begin{eqnarray}
	\rateQ{\A}{\G}=\rateQ{\G}{\A}=\rateQ{\C}{\T}=\rateQ{\T}{\C}&
	\!\!\!\!\!\!=&\!\!\!\!\!\!q\,,
	\label{eqts}
	\\
	\rateQ{\A}{\C}=\rateQ{\C}{\A}=\rateQ{\A}{\T}=\rateQ{\T}{\A}=
\rateQ{\C}{\G}=\rateQ{\G}{\C}=\rateQ{\G}{\T}=\rateQ{\T}{\G}&
	\!\!\!\!\!\!=&\!\!\!\!\!\!p\,.
	\label{eqtv}
	\end{eqnarray}

For this simple case, the single and dinucleotide frequencies 
can be solved analytically in {\em closed form}  
under the two-cluster approximation as described in the Appendix.
Note that since the stationary state does not involve any time scale in itself,
the results only depend on two effective parameters, $q/p$ and $r/p$.
The transition/transversion ratio has been estimated to be  $q/p \approx 3$
previously, based on mammalian pseudogene studies\cite{Go82,blake,Li84,Ya99}.
Combined with the observed 10-fold difference\cite{Batzer,blake} 
in point substitution counts 
between \CpG\ and non-\CpG\ sites 
[which implies $2(r+q) \approx 10 \times
(2p+q)$], we obtain $r/q \approx 20$ for mammals.
The nucleotide frequencies computed according to these rates are 
shown in Table~\ref{tabappx}. They 
compare very well to the results of Monte-Carlo simulation:
the latter performed for an $L=10^8$ system (and averaged over 100
simulations) yielded results that
were within an rms deviation of $10^{-5}$ in the single-nucleotide 
frequencies and $4\cdot10^{-5}$ in the dinucleotide frequencies. 

\begin{table}[ht]
\def\ncA{0.28606}
\def\ncC{0.21394}
\def\ncG{0.21394}
\def\ncT{0.28606}
\def\ncAA{0.07475}
\def\ncAC{0.06119}
\def\ncAG{0.06827}
\def\ncAT{0.08182}
\def\ncCA{0.08052}
\def\ncCC{0.05071}
\def\ncCG{0.01442}
\def\ncCT{0.06827}
\def\ncGA{0.05625}
\def\ncGC{0.04577}
\def\ncGG{0.05071}
\def\ncGT{0.06119}
\def\ncTA{0.07451}
\def\ncTC{0.05625}
\def\ncTG{0.08052}
\def\ncTT{0.07475}
\begin{center}
\begin{tabular}{r|c|cccc}
		$\X$&$f_\X$&$f_{\X\A}$&$f_{\X\C}$&$f_{\X\G}$&$f_{\X\T}$\\
		\hline
	\A&\ncA&\ncAA&\ncAC&\ncAG&\ncAT\\
	\C&\ncC&\ncCA&\ncCC&\ncCG&\ncCT\\
	\G&\ncG&\ncGA&\ncGC&\ncGG&\ncGT\\
	\T&\ncT&\ncTA&\ncTC&\ncTG&\ncTT\\
\end{tabular}
\end{center}
\caption{\label{tabappx}Single and dinucleotide frequencies from 
the {\em two-cluster approximation} 
for  $q/p=3$ and $r/p=20$.
}
\end{table}

Let us examine the results presented in Table~\ref{tabappx}.
First we note 
that the nucleotide composition $f_{\X}$ is skewed towards \A\ and \T,
with a \C+\G\ content of $42\%$, which is 
in general agreement with the typical
average \C+\G\ content in mammals. Note that unlike most existing 
phylogenetic studies\cite{Ya94}, where the mutation rates $\Q$ are {\em tuned}
to reproduce the observed frequencies. In our study, the mutation matrix
$\Q$ by itself would have generated equal \C+\G\ and \A+\T\ content, 
with the anomaly coming solely from the \CpG -methylation-deamination 
process. 
Similar observation was made in \cite{FrZu00} 
based on Monte Carlo simulations.

Next, we analyze the dinucleotide frequencies. It is
useful to focus on the dinucleotide  {\em odds ratio},
$\rho_{\X\Y}\equiv {f_{\X\Y}}/{(f_\X f_\Y)}$,
introduced to indicate whether a specific dinucleotide pair $\X\Y$ is over-
($\rho_{\X\Y}>1$) or under-represented ($\rho_{\X\Y}<1$) with respect to 
neighbor-independent mutation.  
The results derived from the frequencies in Tables~\ref{tabappx} 
are shown in Table \ref{tabappz} (top). 
For comparison, the odds ratio $\rhohat_{\X\Y}$ obtained from 
the {\em observed} single and dinucleotide frequencies 
$\fhat_\alpha$ and $\fhat_{\alpha\beta}$  for a region of 4Mbp 
integenic DNA taken from the Human genome Chromosome 21
(Accession \# NT 011512) is given in Table~\ref{tabappz} (bottom).  
[We will denote each observed
quantity by a ``hat'' throughout the text.]
The rms deviation between the observed ratio $\rhohat_{\alpha\beta}$
and the computed $\rho_{\alpha\beta}$'s  is 0.1. 
We note that the computed ratios capture correctly
the key feature of the observed data, i.e., a strong under-representation
of the \C\G\ dinucleotides,  compensated by the
over-representation of \C\A\ and \T\G.

\begin{table}[ht]

\def\nrAA{0.91}
\def\nrAC{1.00}
\def\nrAG{1.12}
\def\nrAT{1.00}
\def\nrCA{1.32}
\def\nrCC{1.11}
\def\nrCG{0.32}
\def\nrCT{1.12}
\def\nrGA{0.92}
\def\nrGC{1.00}
\def\nrGG{1.11}
\def\nrGT{1.00}
\def\nrTA{0.91}
\def\nrTC{0.92}
\def\nrTG{1.32}
\def\nrTT{0.91}

\def\mrAA{1.10}
\def\mrAC{0.87}
\def\mrAG{1.11}
\def\mrAT{0.91}
\def\mrCA{1.20}
\def\mrCC{1.21}
\def\mrCG{0.20}
\def\mrCT{1.11}
\def\mrGA{0.99}
\def\mrGC{1.05}
\def\mrGG{1.22}
\def\mrGT{0.87}
\def\mrTA{0.80}
\def\mrTC{0.99}
\def\mrTG{1.21}
\def\mrTT{1.10}

\begin{center}
\begin{tabular}{r|cccc}
		$\rho_{\X\Y}$&$\Y=\A$&\C&\G&\T\\
		\hline
	$\X=\A$&\nrAA&\nrAC&\nrAG&\nrAT\\
	     \C&\nrCA&\nrCC&\nrCG&\nrCT\\
 	     \G&\nrGA&\nrGC&\nrGG&\nrGT\\
	     \T&\nrTA&\nrTC&\nrTG&\nrTT\\
            \end{tabular}

\vspace*{2mm}

\begin{tabular}{r|cccc}
		$\rhohat_{\X\Y}$&$\Y=\A$&\C&\G&\T\\
		\hline
	$\X=\A$&\mrAA&\mrAC&\mrAG&\mrAT\\
	     \C&\mrCA&\mrCC&\mrCG&\mrCT\\
 	     \G&\mrGA&\mrGC&\mrGG&\mrGT\\
	     \T&\mrTA&\mrTC&\mrTG&\mrTT\\
            \end{tabular}
\end{center}
\caption{\label{tabappz} Top: Dinucleotide odds ratios of 
the model based on the single and dinucleotide frequencies
in Table~\ref{tabappx}. Bottom: odds ratios obtained for
a region of 4Mbp 
integenic DNA taken from the Human genome Chromosome 21
(Accession \# NT 011512).
}
\end{table}

At this stage of the analysis, 
we cannot expect a perfect match between the observed and 
the computed frequencies in Table~\ref{tabappz}, since we used 
a very rough estimate on the rates $q$ and $r$ in the computation. 
One can of course always tune $q$ and $r$ to match
the computed odds ratios in \C\G\ and \C\A/\T\G. But this is
not sufficient for deducing the underlying mutation rates, 
since the  \CpG\ mutation process introduced also
affects the other dinucleotide counts: As seen in Table~\ref{tabappz} (top),
our model produces 9 other dinucleotides that are over- or under- 
represented by $\sim 10\%$ each. They result 
from secondary mutation since all these 9 pairs are only {\em one} point
mutation away from {\tt CG}, {\tt CA},  or {\tt TG}. Only the four
dinucleotides ({\tt AC}, {\tt AT}, {\tt GC}, {\tt GT}) that
are {\em two} point mutations away 
are not affected by the \CpG\ process.
Note that the magnitude of the change in $\rho_{\X\Y}$ due to  the primary 
mutation, as well as the magnitude and sign of change in $\rho_{\X\Y}$
due to the secondary mutations
could not have been anticipated directly from the model defined by 
Eqs.~(\ref{eqr})--(\ref{eqtv}). 
Thus, the ``backward analysis'' of matching {\em all} 
of the dinucleotide ratios (as well as the single nucleotide 
frequencies) by varying the rates
$q$ and $r$ is a highly nontrivial numerical task. 
Fryxell and Zuckerkandl\cite{FrZu00} attempted to do this by focusing on
a few selected nucleotide frequencies (namely, {\tt CG}, {\tt TG}, 
{\tt TA} and {\tt AT}) which they obtained for different mutation
rates using extensive Monte Carlo simulation.  Their method is 
arbitrary, inaccurate and very time consuming. 
With our approach, we can provide a systematic, accurate, 
and efficient method to perform the backward analysis 
as described below. Our method is implemented in a server
at the same url as provided above, {\tt http://bioinfo.ucsd.edu/dinucleotides} .

\subsection{Backward analysis: Estimation of mutation rates}
\vspace{1mm}

Our next task is to estimate the mutation rates \Q\ and \R\ 
which best ``explain'' the observed data as quantified by 
single and dinucleotide frequencies $\fhat_\alpha$ and 
$\fhat_{\alpha\beta}$. We will adopt the
strategy of maximum likelihood\cite{ml}. Specifically,
we compute  the likelihood ${\cal L}(\Q,\R)$ of
observing the data given our mutation model with the rates
\Q\ and \R: First we note that the probability $P(\vec\alpha)$ of 
observing a sequence $\vec\alpha$ of length $L$ is
$$
P(\vec\alpha) = f_{\X_1\ldots\X_L} \approx
\prod_{i=2}^{L} (f_{\X_{i-1}\X_{i}}/f_{\X_{i}})
$$
according to the two-cluster
approximation (see Eq.~(\ref{approx0}) and the Appendix).
The likelihood of the occurrence of a particular
sequence with  frequencies $\fhat_\X$ and
$\fhat_{\alpha\beta}$ is then given by
$(\prod_{\X\Y} f_{\X\Y}^{L\fhat_{\X\Y}})/(\prod_{\X} f_{\X}^{L\fhat_{\X}})$
yielding the following expression for the log-likelihood,
	\begin{equation}
	\log {\cal L}=L \sum_{\X\Y} {\fhat_{\X\Y}} \log (f_{\X\Y})-
	L \sum_{\X} {\fhat_{\X}}\log(f_{\X}),
	\end{equation}
where $f_\X$ and $f_{\X\Y}$ denote the single- and dinucleotide 
frequencies according 
to our model with rates \Q\ and \R. 

In principle, one can now
search through all of the single-point mutation rates as embodied
in \Q\, and all possible combinations of the neighbor-dependent
mutation processes and their rates as embodied in \R\ to maximize $\L$. 
However, this search space is much too large even given our solution.
More importantly, we do not want to have too many parameters to
over-fit the data. Thus, we allow only a single transversion rate
(which is set to $1$ without loss of generality), 
two transition rates, $\rateQ{\A}{\G}=\rateQ{\T}{\C}$ and
$\rateQ{\C}{\T}=\rateQ{\G}{\A}$ (since a difference in those rates has already
been reported in the literature\cite{PeHa}), and one single
neighbor-dependent mutation process. However, we do not limit the
latter to \CpG\ and search through all of the 48 possibilities 
and their accompanying rates. 
If the maximum-likelihood solution found  is 
still not satisfactory, then an additional neighbor-dependent process 
may be included to further improve the result.

To illustrate the backward analysis, we feed in the single and
 dinucleotide counts of the above mentioned intergenic region of
Human Chromosome 21. The best neighbor-dependent process 
found by our program is the known 
process $\CpG \to {\tt CpA}$ or $\CpG \to {\tt TpG}$.  
However, the transition rates corresponding to the maximum
likelihood solution are at the boundary of the region searched.
They are even smaller than the transversion rate $p=1$, indicating
that there is something wrong, most likely a sign that
the data contained some feature(s) not anticipated by our
mutation model. In order to get a hint at the source of the problem,
we repeated the search allowing for 
an additional neighbor-dependent mutation process
which would bring the transition rates to the expected regime.
 The program found the additional processes
${\tt ApC} \to {\tt ApA}$ or ${\tt GpT} \to {\tt TpT}$; with
these processes, the maximum-likelihood solution for the mutation 
rates became more reasonable, with
$\rateQ{\A}{\G}=\rateQ{\T}{\C}=3.10\, p$,
$\rateQ{\C}{\T}=\rateQ{\G}{\A}=3.78\, p$,
$\rateR{\C\G}{\C\A}=\rateR{\C\G}{\T\G}=43.02\, p$, and 
$\rateR{\A\C}{\A\A}=\rateR{\G\T}{\T\T}=4.35\, p$.
Since we are not aware of any biological or biochemical evidence
for the additional mutation processes
${\tt ApC} \to {\tt ApA}$ or ${\tt GpT} \to {\tt TpT}$,
we interpret this result merely as an indication that the over-abundance
of \A\A\ and \T\T\ in the observed data (see Table~\ref{tabappz})
makes it difficult to ``explain'' by the model with only the 
\CpG\ mutation process.  (From the top panel of Table~\ref{tabappz},
it is clear that the \CpG\ process only {\em reduces} the odds ratio
of \A\A\ and \T\T, while the observed trend is the opposite.)
\begin{figure}
\begin{center}
\setlength{\unitlength}{8.0cm}
\begin{picture}(1,0.65)(0,0)
\put(0,-0.1){\includegraphics[width=\unitlength]{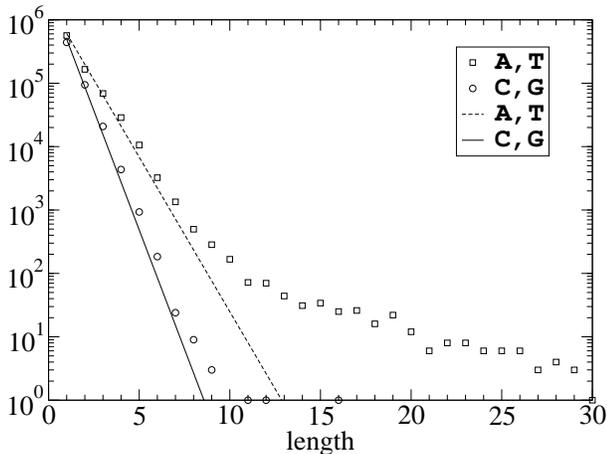}}
\end{picture}
\end{center}
\caption{\label{fig}The number of homonucleotide subsequences 
of a given length in the above mentioned sample of 4Mbp intergenic DNA 
taken from the Human Chromosome 21.
The straight lines represent the expected distributions 
of such subsequences according to
their observed single-nucleotide 
frequencies of bases and assuming no correlations between neighboring
bases. }
\end{figure}

What might be the source of this over-ab\-und\-ance?
From an inspection of the length distribution of the 
homonucleotide subsequences (Fig.~\ref{fig}), one sees 
an conspicuous over-ab\-un\-dance of long runs of \A's or \T's.
These long runs, while low in numbers, 
will clearly bias the dinucleotide counts.
However, their occurrences are thought
to arise from sequence-specific insertion processes such as
polymerase slippage and have nothing to do with neighbor-dependent
mutation being studied here. In order to perform the 
 backward analysis properly, it is therefore necessary to first
{\em filter out} the effect of such processes. The odds ratio
$\rhohat^{\rm (filter)}$
obtained after removing all homonucleotide subsequences of length 
four or more  are
presented in Table~\ref{tabapphs} (top).  Comparing it to 
Table~\ref{tabappz} (bottom), we see substantial ($> 15\%$) decreases 
in the counts for \A\A, \T\T, \C\C, and \G\G\ as expected, 
as well as  $> 10\%$ increases in \A\T\ and \T\A. (The single 
nucleotide frequencies are however not affected much by the filter.)
Applying the backward analysis on the filtered data with a single
neighbor-dependent process, we recover the known \CpG\ process
again, but this time with the rates
$\rateQ{\A}{\G}=\rateQ{\T}{\C}=1.75\, p$,
$\rateQ{\C}{\T}=\rateQ{\G}{\A}=2.33\, p$, and
$\rateR{\C\G}{\C\A}=\rateR{\C\G}{\T\G}=22.17\, p$,
which are all within the known range given above. [Similar
results were obtained when we filtered out subsequences of length 5 
or longer with at least 80\% of the same nucleotide. The rates found
were
$\rateQ{\A}{\G}=\rateQ{\T}{\C}=1.66\, p$,
$\rateQ{\C}{\T}=\rateQ{\G}{\A}=2.03\, p$, and
$\rateR{\C\G}{\C\A}=\rateR{\C\G}{\T\G}=17.80\, p$.]
The predicted odds ratio $\rho^*$ corresponding to these maximum-likelihood
rates are
presented in Table~\ref{tabapphs} (bottom). The rms deviation 
between the filtered data and the maximum-likelihood prediction 
is $0.05$\,.

\begin{table}[ht]

\def\nrAA{0.91}
\def\nrAC{0.92}
\def\nrAG{1.18}
\def\nrAT{1.02}
\def\nrCA{1.26}
\def\nrCC{1.05}
\def\nrCG{0.23}
\def\nrCT{1.18}
\def\nrGA{1.02}
\def\nrGC{1.02}
\def\nrGG{1.05}
\def\nrGT{0.92}
\def\nrTA{0.90}
\def\nrTC{1.02}
\def\nrTG{1.26}
\def\nrTT{0.91}

\def\mrAA{0.93}
\def\mrAC{1.00}
\def\mrAG{1.11}
\def\mrAT{1.00}
\def\mrCA{1.30}
\def\mrCC{1.09}
\def\mrCG{0.24}
\def\mrCT{1.11}
\def\mrGA{0.94}
\def\mrGC{1.00}
\def\mrGG{1.09}
\def\mrGT{1.00}
\def\mrTA{0.91}
\def\mrTC{0.94}
\def\mrTG{1.30}
\def\mrTT{0.93}

\begin{center}
\begin{tabular}{r|cccc}
		$\rhohat^{\rm (filter)}_{\X\Y}$&$\Y=\A$&\C&\G&\T\\
		\hline
	$\X=\A$&\nrAA&\nrAC&\nrAG&\nrAT\\
	     \C&\nrCA&\nrCC&\nrCG&\nrCT\\
 	     \G&\nrGA&\nrGC&\nrGG&\nrGT\\
	     \T&\nrTA&\nrTC&\nrTG&\nrTT\\
            \end{tabular}

\vspace*{2mm}

\begin{tabular}{r|cccc}
		${\rho}^{*\phantom{()}}_{\X\Y}$&$\Y=\A$&\C&\G&\T\\
		\hline
	$\,\,\X=\A$&\mrAA&\mrAC&\mrAG&\mrAT\\
	     \C&\mrCA&\mrCC&\mrCG&\mrCT\\
 	     \G&\mrGA&\mrGC&\mrGG&\mrGT\\
	     \T&\mrTA&\mrTC&\mrTG&\mrTT\\
            \end{tabular}
\end{center}
\caption{\label{tabapphs}Top: Observed dinucleotide odds ratios 
of intergenic DNA taken from the Human genome Chromosome 21 
(Accession \# NT 011512) with all homonucleotide subsequences 
of four or more bases filtered out.
Bottom:  The odds ratio $\rho^*$ corresponding to the maximum-likelihood
solution of the backward analysis.
The rms deviation between the two odds ratios is $0.05$\,.
}
\end{table}
\section{Discussion and Outlook}
\vspace{1mm}

We have introduced a model of neighbor-dependent mutation and
developed methods to study these processes which are believed to play
an important role in the evolution of genomic sequences. We see that
the base composition and the dinucleotide frequencies are strongly 
influenced by neighbor-dependent mutation. This may contribute to 
the (surprising) success of  phylogenetic analysis based on dinucleotide 
counts\cite{KM97,Karetal97}, since the pattern of dinucleotide frequencies reflect
the mutation mechanisms and rates which should be highly conserved 
throughout evolution. 
A similar mutation model was investigated in Ref. \cite{FrZu00} using
Monte Carlo simulations. 
In that work, the backward analysis
required a number arbitrary assumptions 
and large amounts of computer time.
Our work represents a systematic
approach to these studies. The analysis tools we provide enable the users
to examine a large number of mutation processes and rates in detail
and over a large number of genomes, to look for previously unknown mutation 
processes and  track them across the different kingdoms of life.

Of course, the mutation model (i.e., Eqs.~(\ref{rulei}) and (\ref{ruleii}))
and the assumption 
of stationarity for observed sequences need to be verified quantitatively.
But the present method is in fact not limited to the study
of the stationary state, and can be straightforwardly extended to 
describe the evolution  towards stationarity\cite{ArHw}. This can be used
to extend the traditional phylogenetic analysis\cite{Ya94} to include neighbor-dependent effects, and should be particularly
useful for the vertebrates which are dominated by
the \CpG-methylation-deamination process. 
Finally we note that a more accurate description of the evolution 
of non-coding genomic DNA sequences will be of value to various
sequence analysis applications, particularly those involving mammalian
genomes. For instance, in the comparative genomics approach 
to gene and DNA motif finding, one needs to evaluate the probability 
of spurious matches
due to shared common ancestry in the non-coding regions\cite{Frazer}.  
Accurate evaluation of such
probabilities depend critically on having realistic models of sequence evolution,
which this study contributes towards.

\section{Acknowledgment}
\vspace{1mm}
The authors are grateful to the hospitality of the Institute for Theoretical 
Physics (Santa Barbara) where this work was initiated.
 PFA and TH are supported by the NSF through grant DMR-9971456.
PFA additionally acknowledges the financial support by a DFG fellowship, and
CBB and TH
by  functional genomics innovation awards from the Burroughs Wellcome Fund.


\appendix

\section{The Two-Cluster Approximation}
\label{secexa}

\vspace{1mm}
To illustrate our method, we consider a specific example
of  the neighbor-dependent mutation process, the \CpG-methy\-lation-deamination
process mentioned in the introduction.
A similar model has also been studied using Monte-Carlo simulations in 
Ref. \cite{FrZu00}.
Given the mutation processes defined by Eqs.~(\ref{eqr}), (\ref{eqts}) and
(\ref{eqtv}), the time evolution of the
dinucleotide frequency $f_{\X\Y}$ can be expressed in terms of the function
$f_{\X\Y}$ and the trinucleotide frequency $f_{\X\Y\Z}$ only.  It will be convenient 
to take the continuum time limit, and turn the discrete dynamics described in
Sec.~2 into differential equations\cite{Gardinger}. 
For example, the
evolution of $f_{\A\A}$  is given by 
	\begin{eqnarray}
	\frac{\partial}{\partial t}f_{\A\A}
	\!\!\!\!\!&=&\!\!\!\!
	p f_{\underline{\C}\A} + q f_{\underline{\G}\A} +p f_{\underline{\T}\A} 
	-(2p+q) f_{\underline{\A}\A}
	\nonumber\\
	&&\!\!\!\!\!+p f_{\A\underline{\C}} +q f_{\A\underline{\G}} + p f_{\A\underline{\T}}
	-(2p+q) f_{\A\underline{\A}} +r f_{\underline{\C\G}\A}
	\label{caasample}
	\end{eqnarray}
where all terms on the right hand side correspond to processes either creating (positive
signs) or destroying (negative signs) an \A\A\ pair. In Eq.~(\ref{caasample}),
the first four terms
result from point mutations on the first site, the second four from point
mutations on the second site, as indicated by the underlined letters.  The
last term proportional to $r$ stems from a $\G$ turning into an $\A$ due to a
$\C\G\rightarrow\C\A$ process on the two sites shifted by one to the left.
There are $16$ such equations for the $16$ dinucleotide frequencies.

As one can see in Eq.~(\ref{caasample}), the time evolution of the
two-point functions depends in general on the three-point functions
$f_{\X\Y\Z}$ which in turn will depend on the four-point functions.
To truncate this hierarchy of equations, we apply a standard closure 
approximation used in non-linear dynamics\cite{bAKo92}
	\begin{equation}
	f_{\X\Y\Z}=\frac{f_{\X\Y}f_{\Y\Z}}{f_{\Y}}
	\label{approx}
	\end{equation}
i.e. we approximate the probability of finding three letters $\X\Y\Z$
by the probability of finding a pair $\X\Y$ multiplied by
the {\em conditional} probability of finding a pair $\Y\Z$, given that the middle
base is $\Y$; the latter is expressed according to Bayes's rule.
The $L$-point function subsequently takes the form
	\begin{equation}
	f_{\X_1\dots\X_L}\approx\prod_{i=2}^{L} \frac{f_{\X_{i-1}\X_{i}}}{f_{\X_{i}}}
	\label{approxL}
	\end{equation}
The single nucleotide frequencies called for in Eqs.~(\ref{approx}) and (\ref{approxL})
can be expressed in terms of the dinucleotide frequencies as
	\begin{equation}
	f_{\X}=\sum_{\Y} f_{\X\Y}=\sum_{\Y} f_{\Y\X}.
	\label{oneclu}
	\end{equation}
Thus, the right hand side of Eq.~(\ref{caasample}) can be written as a 
function of the $f_{\X\Y}$'s only. We can generate an equation similar to
(\ref{caasample}) for each of the 16 $f_{\X\Y}$'s. Applying 
the approximation (\ref{approx}) and the identity (\ref{oneclu}),
we then obtain a {\em closed} system of  coupled, nonlinear 
differential equations
	\begin{equation}
	\frac{\partial}{\partial t} f_{\X\Y}=
	{\cal G}_{\X\Y} (f_{\A\A},f_{\A\C},\dots,f_{\T\T})
	\label{eqsys}
	\end{equation}
with 16 ${\cal G}_{\X\Y}$'s.
In the stationary state, the functions $f_{\X\Y}(t)$ are independent of $t$
and hence their derivative with respect to $t$ vanishes.
To calculate the stationary $f_{\X\Y}$'s,  we therefore have to
solve the set of 16 (quadratic) equations, 
${\cal G}_{\X\Y} (f_{\A\A},f_{\A\C},\dots,f_{\T\T})=0$,
whose solution is
straightforwardly obtained with the help of Mathematica and given below:

It is convenient to express the results in term of the parameter
	\begin{equation}
	\Delta=
	\frac{(3p+q)r}{16(p+q)(3p+q)+4(7p+3q)r}.
	\end{equation}
The single nucleotide frequencies are
	\begin{equation}
	f_\A=f_\T=\frac14+\frac\Delta2
	\;,\quad
	f_\C=f_\G=\frac14-\frac\Delta2.
	\end{equation}
Since $\Delta$ is an increasing function of $r$ with $\Delta(r=0)=0$, 
we see that $f_\A=f_\T > 1/4$ for all positive $r$'s.

The dinucleotide frequencies are most succinctly expressed in terms
of the auxiliary functions $\hat{f}_{\X\Y} \equiv f_{\X\Y} - f_\X f_\Y$.
The results are:
	\begin{eqnarray}
	\hat{f}_{\C\A} =
	\frac{\Delta(1+\Delta)}{4}\,,\qquad
	\hat{f}_{\C\G} =
	-\frac{r(1-2\Delta)^2-16(p+q)\Delta}{16r}\\
	\hat{f}_{\C\C} =
	-\frac{(2\Delta-1)(4r\Delta^2+8(2p+2q+r)\Delta-r)}{32r(\Delta-1)}.
	\end{eqnarray}
Additionally, we have 
	\begin{equation}
	\hat{f}_{\A\C}=\hat{f}_{\A\T}=\hat{f}_{\G\C}=\hat{f}_{\G\T}=0
	\end{equation}
since these four dinucleotides are {\em two} mutations away from 
the three primary pairs above, as motivated already in Sec.~\ref{secres}.

The remaining 9 frequencies can now be obtained simply by exploiting
the relation $\sum_\X \hat{f}_{\X\Y}=0=\sum_\X \hat{f}_{\Y\X}$ which follows
from Eq.~(\ref{oneclu}), and the reverse complementarity symmetry
$\hat{f}_{\X\Y} = \hat{f}_{\bY\bX}$. The results are:
	$$
	\hat{f}_{\C\T}=-(\hat{f}_{\C\A}+\hat{f}_{\C\C}+\hat{f}_{\C\G})
	\;,\quad
	\hat{f}_{\T\C}=-\hat{f}_{\C\C}
	\;,\quad
	\hat{f}_{\T\T}=-\hat{f}_{\C\T},
	$$
	$$
	\hat{f}_{\A\A}=\hat{f}_{\T\T}
	\;,\quad\!
	\hat{f}_{\A\G}=\hat{f}_{\C\T}
	\;,\quad\!
	\hat{f}_{\G\A}=\hat{f}_{\T\C}
	\;,\quad\!
	\hat{f}_{\G\G}=\hat{f}_{\C\C}
	\;,\quad\!
	\hat{f}_{\T\G}=\hat{f}_{\C\A},
	$$
	$$
	\hat{f}_{\T\A}=-(\hat{f}_{\A\A}+\hat{f}_{\C\A}+\hat{f}_{\G\A}).
  	$$

As shown already in Sec.~\ref{secres},
this approximation gives very accurate results
when  compared to Monte-Carlo simulations of the same system.
This is due to the very short-ranged correlation induced by the 
mutation process and will not be elaborated here.

\end{document}